\documentclass{IEEEcsmag}

\usepackage[colorlinks,urlcolor=blue,linkcolor=blue,citecolor=blue]{hyperref}
\expandafter\def\expandafter\UrlBreaks\expandafter{\UrlBreaks\do\/\do\*\do\-\do\~\do\'\do\"\do\-}
\usepackage{upmath,color}
\usepackage{amssymb}
\usepackage{comment}
\usepackage[linesnumbered,ruled,vlined]{algorithm2e}
\usepackage{xcolor}
\usepackage{soul}
\usepackage{adjustbox}
\usepackage{pifont}
\newcommand{\cmark}{\ding{51}}%
\newcommand{\xmark}{\ding{55}}%

\jvol{XX}
\jnum{XX}
\paper{8}
\jmonth{Month}
\jname{}
\jtitle{Adversarial Explainability: Utilizing Explainable Machine Learning in Bypassing IoT Botnet Detection Systems}
\pubyear{2023}

\begin{document}

\sptitle{Regular Article}

\title{Adversarial Explainability: Utilizing Explainable Machine Learning in Bypassing IoT Botnet Detection Systems}

\author{Mohammed M. Alani}
\affil{Toronto Metropolitan University, Toronto, Canada}

\author{Atefeh Mashatan}
\affil{Toronto Metropolitan University, Toronto, Canada}

\author{Ali Miri}
\affil{Toronto Metropolitan University, Toronto, Canada}


\begin{abstract}\looseness-1Botnet detection based on machine learning have witnessed significant leaps in recent years, with the availability of large and reliable datasets that are extracted from real-life scenarios. Consequently, adversarial attacks on machine learning-based cybersecurity systems are posing a significant threat to the practicality of these solutions. In this paper, we introduce a novel attack that utilizes machine learning model's explainability in evading detection by botnet detection systems. The proposed attack utilizes information obtained from model's explainability to build adversarial samples that can evade detection in a blackbox setting. The proposed attack was tested on a trained IoT botnet detection systems and was capable of bypassing the botnet detection with 0\% detection by altering one feature only to generate the adversarial samples.
\end{abstract}

\maketitle
\footnote{This work has been submitted to the IEEE for possible publication. Copyright may be transferred without notice, after which this version may no longer be accessible}
\section{INTRODUCTION}
Machine-learning (ML) has a received lot of attention in various research areas in recent years. According to \cite{ref1}, the ML market revenue in 2023 is projected to reach \$165 billion, with projected value of \$2,028 billion in five years. Cybersecurity is a field where ML-based solutions have shown noticeable potential, especially with the recent increase in cyber attacks. ML-based cyber attack detection systems that are built for Cyber-Physical Systems (CPS) have also gained popularity, despite the limited resources available in these systems.

With the large amounts of security-related data and logs being generated from various vantage points within the industrial, enterprise, and personal networks, machine learning presents itself as a capable solution that can ingest these large amounts of data and produce useful predictions with continuously improving accuracy. With the rapid growth in adopting Internet-of-Things (IoT) devices in different areas of our lives, security threats to these devices grow in parallel. Within different contexts, such as industrial, health, smart city, etc., CPSs generate heaps of network and system logs in a rapidly growing fashion. In addition, the use of advanced obfuscation and evasion techniques by adversaries makes legacy and signature-based detection systems incapable of keeping up with the modern-day security challenges. The intersection of cybersecurity and ML research has witnessed many significant developments in various areas of cybersecurity as seen in \cite{ref2,ref4}. These applications mostly reside within different areas of ``detection'', such as intrusion detection, malware detection, phishing detection, vulnerability detection, and spoofing detection. Machine learning solutions were also introduced to support forensics evidence identification and classification. The rapid shift towards CPS, cloud-based systems, and work-from-home has made the attack surface and ever changing one. This has increased the reliance on ML-based security solutions as an important tool capable of supporting defense efforts. Botnets have specifically grown in popularity among malicious actors in the recent years. Botnets are defined as a network of internet-connected devices infected by a special type of malware that enables an attacker (the bot herder) to control these devices and use them to attack other targets~\cite{ref23}. ML-based botnet detection is one of the areas that have witnessed an increase in research in recent years. Different ML-based techniques were used for that purpose as found in \cite{ref24}, and \cite{ref25}.

ML-based systems themselves were subject to many attacks throughout the years. Some of these attacks target the algorithms used with the aim of altering the output of the algorithm and push it to misbehave. Other attacks targeted the data used in training these systems trying to infer the data to impact its privacy or in an attempt to re-create a similar system using the same dataset~\cite{ref10}.

Many ML-based botnet detection systems operate in hostile environments that are rich with adversaries that want to see these systems fail. Hence, these systems are considered an important target for adversaries. Adversaries aim at evading detection, or bringing down the whole system at once \cite{ref10}. Most of these attacks focus on generating adversarial samples that are capable of bypassing detection through manipulating certain features within these samples \cite{ref13}.

Another current trend in ML-based applications is the use of explainable ML. Explainable machine learning focuses on explaining how machine learning models make their predictions. Explainability helps in finding the impact of each feature of the prediction decision. Such feat would significantly increase trust in these systems by making them more understandable by humans. Significant research has been conducted to facilitate the explainability of the ML-models used in cybersecurity to increase trust, and improve transparency of these models. Recent years have witnessed an increase in the use of explainable machine learning in cybersecurity solutions to ensure that the decisions made by these systems originate from a an explainable system rather than a ``blackbox'' \cite{ref14}.

In this paper, we present a novel adversarial attack on machine learning-based botnet detection solutions, that relies on model's explainability. The proposed attack uses explainability to identify the most effective features in the classifier. Then, these features are used to create adversarial samples that lead to evading detection. Such technique helps the attacker in identifying the changes that need to be made to the attack technique to help it become indetectable by machine learning-based detection systems.

The next section introduces a review of related works, and discusses several of the most recent adversarial machine learning attacks. Section 3 
introduces the concept of the proposed attack along with detailed explanation of the algorithm used to generate the adversarial samples. Section 4 
shows the detailed steps of the experiments conducted to test the efficacy of the proposed attack, along with the obtained results. The discussions of the obtained results are provided in Section 5 
along with a comparative analysis of the proposed attack with related works. The conclusions and future research directions are discussed in Section 6.

\section{RELATED WORKS} 
\label{sec:related}

Most of relevant previous works were focused on the generation of adversarial samples. The generation of these samples relies on different techniques such as Generative Adversarial Networks (GANs), and heuristic techniques. In this section we will review several of the most recent attacks that aim at evading detection in general, with a focus on evading intrusion botnet and intrusion detection.

Lin et al. presented, in 2018, a GAN-based approach that targets machine learning-based intrusion detection systems, named IDSGAN \cite{ref15}. the proposed system follows a black-box approach assuming that the attacker has no prior knowledge of the IDS model, and leverages a generator that can transform original malicious samples into adversarial attack samples. The proposed system was tested using NSL-KDD dataset and proved to be effective in reducing the accuracy of the detection process from 79.12\% for denial of service attacks, to only 0.61\%.

Ayub et al. presented, in 2020, an adversarial attack that aims to evade machine learning-based intrusion detection systems \cite{ref12}. The attack utilizes Jacobian Saliency Map Attack (JSMA) to find the most effective features that would impact the accuracy of the model significantly. The proposed attack was tested on an IDS system trained using CIC-IDS-2017 dataset. Testing showed that the attack succeeded in reducing the accuracy of the detection process from 99.5\% to 78.09\%.

Sheatsley et al. published, in 2022, a study examining the impact of constrained vs. unconstrained environments on adversarial example generation in ML-based intrusion detection systems \cite{ref16}. The study also presented Adaptive-JSMA, and augmented version of JSMA that can be utilized in constrained domains. the study found that constrained domains, such as network traffic manipulation to evade detection, are not significantly more robust to adversarial attacks when compared to unconstrained domains, such as image recognition. The study found that as little as five random features can cause significant degradation in the detection accuracy.

In 2023, Mohammadian et al. presented an adversarial attack designed to evade deep learning-based intrusion detection system \cite{ref11}. The proposed attack uses JSMA to find the best group of features, with different features and perturbation magnitude, to generate adversarial samples that are capable of bypassing detection. The proposed work was tested on three different datasets and proved the proposal's ability to deviate 18\% of the samples in CIC-IDS-2017 dataset by manipulating three features only. The presented attack utilizes a similar method to that used in \cite{ref12}.

Debicha et al. presented a botnet detection evasion attack based on generating adversarial perturbations based on statistical feature manipulation~\cite{ref26}. The study focused on the practicality of such attacks and whether they could be implementable in real life or not. Test conducted using CTU-13, and CSE-CIC-IDS2018 datasets showed that the proposed attack was detected with accuracy of 5.3\% to 22.2\% for different botnet families. The study also presented a method of detection that would improve immunity of ML-based detection systems against the proposed attack. 

Further details about adversarial attacks against ML-based botnet and intrusion detection systems can be found in \cite{ref18}.

\section{ADVERSARIAL EXPLAINABILITY} 
\label{sec:proposed}
Most recent explainability techniques, such as Local Interpretable Model-agnostic Explanations (LIME), and Shapley Additive Explanations (SHAP), are focused on analyzing the ML model's behavior and identifying the impact of each individual feature on the prediction outcome of the model \cite{ref22}. This goes beyond classical techniques such as feature importance, by identifying the extent of the impact of the value of feature on the value of the prediction resulting from the model.

SHAP, which was introduced in 2017 \cite{ref19}, is a model-agnostic technique of explainability based on game theory. The technique calculates the impact of each feature by comparing the model's performance with the feature to the model's performance without the feature. This does not only determine the importance of the feature in the prediction process, but also identifies the specific impact of higher and lower values of this feature on the prediction value.

Explainability techniques such as SHAP can help adversaries address the an important gap in the adversarial machine learning field, which is identifying the most effective features to produce impactful perturbations that could effectively impact the model's accuracy. In fact, this extends beyond identifying the most effective features to the identification of the changes required in the value of these features to produce adversarial samples.

\begin{figure*}
    \centering
    \includegraphics[width=0.65\textwidth]{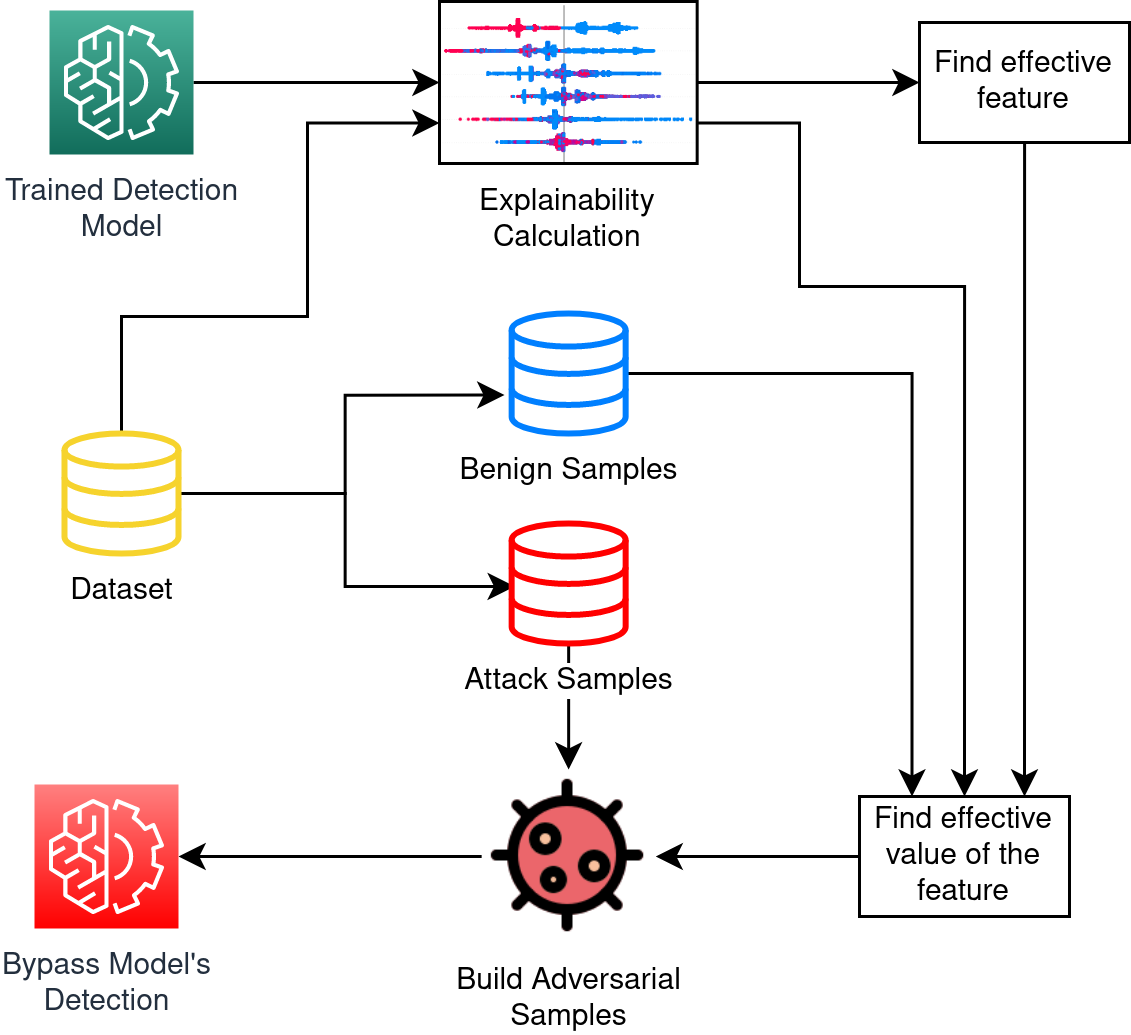}
    \caption{An overview of Adversarial Explainability Attack}
    \label{fig:advex}
\end{figure*}

Figure \ref{fig:advex} shows an overview of the proposed novel attack concept. The proposed attack starts by applying an explainability technique, such as SHAP, to a trained botnet detection model. This would produce detailed measures of the impact of the change of the values of each feature on the prediction result. In the next step of the proposed attack, these explainability values are used in selecting the feature (or features) that cause the highest possible impact on the prediction decision. Once this feature is identified, the search for the most impactful value of this feature starts. The attacker then search through the generated explainability values of that particular feature to find a benign sample where the feature has most significantly contributed to the prediction. The attacker then uses the value of this feature from this sample as the value to be used in generating the adversarial samples. The identified perturbation can then be used in altering the malicious samples to help them get misclassified as ``benign'' ones.

\begin{algorithm} \label{algo:advex}
    \SetAlgoLined
    \DontPrintSemicolon
    \KwIn{\textbf{M} trained model, \textbf{D} dataset}
    \KwOut{\textbf{D\textsubscript{A}} Adversarial samples generated from malicious samples.}
    \BlankLine
    $S \gets SHAP(M,D)$ \tcp*[f]{Calculate SHAP Values}\;
    $F \gets max(|S|)$ \tcp*[f]{Find the location of feature with highest impact}\;
    \ForEach{$d$ in $D$}{
    \If{$d$ is $benign$}{
        $D_B \gets d$ \tcp*[f]{Place all benign samples in $D_B$}\;}
    \Else{
        $D_M \gets d$ \tcp*[f]{Place all malicious samples in $D_M$}\;}
    }
    $\lambda = 0$ \tcp*[f]{Initialize location of the benign sample with the most effective feature value}\;
    $\sigma = S(\lambda,F)$ \tcp*[f]{Initialize the most effective feature impact}\;
    \For{$i = 1$ to $len(S)$}{
        \If{$S(i,F) < \sigma$ AND $D(i,label) = 0$}{
            $\sigma = S(i,F)$\;}    
            $\lambda = i$\;
    }
    $\epsilon = D(\lambda,F)$ \tcp*[f]{The most effective value of the feature F is obtained}\; 
    \For{$i=0$ to $len(D_M)$}{
        $D_A(i) \gets D_M(i)$\;
        $D_A(i,F) \gets \epsilon$\;
        }
    \caption{Generation of adversarial samples}
\end{algorithm}

The detailed steps of generating adversarial samples are shown in Algorithm \ref{algo:advex}.

As shown in the algorithm, the process starts by calculating SHAP values ($S$) of the trained model ($M$), using a dataset ($D$). The SHAP values obtained as a result of these calculations are a two dimensional array. Each row in that array represents the impact of each feature on the prediction decision. If $D$ used in this calculation includes $x$ samples, of $y$ features, the resulting $S$ will have the dimensions $x\times y$. Within this array, $S(i,j)$ represent the impact of feature number $j$ in sample number $i$ on the prediction process. 

Once $S$ is obtained, the location of feature with the highest impact ($F$) is identified by calculating the average SHAP values of each feature throughout all samples. This would generate a \textit{summary} SHAP value where one value representing the impact, or importance, of one feature. In the following step, subsets of the dataset with all of the benign samples ($D_B$), and all of the malicious samples ($D_M$) are created. 

Starting at step 11, the search for the best perturbation value using epxlainability starts. This is done by locating the benign sample where the feature $F$ had the strongest impact (i.e., the highest SHAP value) that pushed the prediction to ``benign''. In this step, a pointer to the location of the sample where the identified feature $F$ has the strongest impact is initialized as $\lambda$. The SHAP value of the most effective feature at the location $\lambda$ is also initialized as $\sigma$.

In steps 13 to 18, the attacker searches for the sample where the SHAP value of feature $F$ exists. Note that the condition in step 14 is smaller than (<). The reason behind this, is that usually, ``benign'' behavior is labelled as class 0, while malicious behvior is labelled class 1. Hence, the attacker is actually searching for the smallest SHAP value (usually a large magnitude negative value) that would give the prediction the biggest push towards 0 (benign) prediction.

Once the location of the lowest SHAP value for feature $F$ is identified, in $\lambda$, the actual feature value is obtained from the dataset using $D(\lambda,F)$, where $\lambda$ identifies the row, and $F$ identifies the column. The obtained value, $\epsilon$, is then assigned to the feature in all of the attack samples extracted earlier to turn them into adversarial samples.

After creating the adversarial samples, they can be used as input to the trained model to test the impact of the perturbations applied to these malicious samples, and check if they result in diverting the prediction to classify these adversarial samples as ``benign'' samples.

\section{EXPERIEMENTS AND RESULTS} \label{sec:experiments}

\subsection{Experiment Design}
To verify and validate the proposed attack, the following experiments were conducted:
\begin{enumerate}
    \item Recreate and test a previously trained botnet detection system that has demonstrated high accuracy.
    \item Calculate SHAP values for the trained model.
    \item Generate adversarial samples using Algorithm \ref{algo:advex}.
    \item Test the trained system using the original dataset and the adversarial samples to compare the results.
\end{enumerate}

\subsection{Implementation Environment}
The dataset preprocessing, and all experiments were performed in an environment with the following specifications:
\begin{itemize}
    \item Processor: AMD Ryzen 9 5950x 16-cores
    \item GPU: NVidia RTX3060Ti - 8GB VRAM
    \item RAM: 128GB
    \item Operating System: Ubuntu 22.04.2
    \item Python: 3.10.10
    \item SciKit Learn: 1.2.2
    \item SHAP: 0.41.0
    \item Matplotlib: 3.7.1
\end{itemize}

\subsection{Performance Metrics}
According to \cite{ref28}, the basic performance metrics for binary classifiers are listed below:
\begin{enumerate}
    \item True positive (TP): The number of tests samples predicted as 1, and their actual class is 1.
    \item True negative (TN): The number of tests samples predicted as 0, and their actual class is 0.
    \item False positive (FP): The number of tests samples predicted as 1, and their actual class is 0.
    \item False negative (FN): The number of tests samples predicted as 0, and their actual class is 1.
\end{enumerate}

These four building blocks can be combined to generate the following metrics:
\begin{enumerate}
    \item Accuracy: measures the ratio of correct predictions using the following equation:
    \begin{equation}
        Accuracy = \frac{TP + TN }{TP + TN + FP + FN}
    \end{equation}
    \item Precision: measures the ratio of the accuracy of positive predictions using the following equation:
    \begin{equation}
        Precision = \frac{TP}{TP + FP}
    \end{equation}
    \item Recall: measures the ratio of positive instances correctly detected by the classifier using the following equation:
    \begin{equation}
        Recall = \frac{TP}{TP + FN}
    \end{equation}
    \item $F_1$ score: measures the harmonic mean of precision and recall using the following equation:
    \begin{equation}
        F_1 Score = 2*\frac{\frac{TP}{TP + FN} * \frac{TP}{TP + FP}}{\frac{TP}{TP + FN} + \frac{TP}{TP + FP}}
    \end{equation}
\end{enumerate}

\subsection{Dataset and Preprocessing}
The botnet detection system used in our experiments to demonstrate the attack was created to replicate the system introduced in \cite{ref23}. Therefore, the dataset used to train, test, and create the adversarial samples was the same dataset used in \cite{ref23}.

The dataset was introduced in \cite{ref27}. The original dataset was comprised of 42 raw network packet (pcap) files capturing different attacks within an IoT environment containing realistic IoT devices such as smart lights and cameras. Following the steps of the original experiment in \cite{ref23}, we extracted the selected features from botnet attack packets, and benign traffic packet while excluding non-botnet attacks. Following the preprocessing steps of the original experiment resulted in a dataset of 919,920 samples, including 691,669 benign samples, and 228,251 malicious samples (75\% benign, 25\% malicious). Each sample includes 7 features extracted from one network packet. The selected features are listed in Table \ref{tab:features}.

\begin{table}[h]
    \centering 
    \caption{Selected features and their description}
    \begin{tabular}{lp{4cm}}
        \hline
        \textbf{Feature} & \textbf{Description} \\
        \hline
        \texttt{frame.len} & Number of bytes within a frame\\
        \texttt{udp.dstport} & Destination port number in a UDP segment \\
        \texttt{ip.flags} & The value of the flags field in the IP packet header\\
        \texttt{tcp.dstport} & Destination port number in a TCP segment \\
        \texttt{ip.ttl} & TTL field value in the IP packet header\\
        \texttt{udp.srcport} & Source port number in a UDP segment\\
        \texttt{ip.len} & Number of bytes in an IP packet, as taken from the IP packet header\\
        \hline
    \end{tabular} \label{tab:features}
\end{table}

\subsection{Botnet detection evasion} \label{sec:botnet}
Once the dataset was preprocessed, we created an extreme gradient boost (XGB) binary classifier to detect botnet traffic. The dataset was randomly split into 75\% training subset, and 25\% testing subset. The random split was performed implemented was stratified to ensure that the percentage of each class is maintained in the training and testing subsets.

The confusion matrix of the testing results are shown in Table \ref{tab:org-results}. In the table, the macro average is a simple averaging of the performance metrics by adding them and dividing by two. A weighted average represents the average generated by multiplying each performance metric of each class by the percentage of this class from the total number of samples~\cite{ref28}. As shown in the table, the accuracy stands at 99.77\%, while the macro average $F_1$ score was 0.9969. This indicates that the trained botnet detection system performs very well.

\begin{table}
    \centering
    \caption{Testing results before creating advesarial samples}
    \begin{tabular}{rlll}
        \hline
        & \textbf{Precision} & \textbf{Recall} & \textbf{$F_1$ Score}\\
        \hline
        \textbf{0} & 0.9990 & 0.9979 & 0.9985 \\
        \textbf{1} & 0.9938 & 0.9971 & 0.9954 \\
        \textbf{Accuracy} &  &  & 0.9977 \\
        \textbf{Macro average} & 0.9964 & 0.9975 & 0.9969 \\
        \textbf{Weighted average} & 0.9977 & 0.9977 & 0.9977 \\
        \hline
    \end{tabular} \label{tab:org-results}
\end{table}

\begin{figure}
    \centering
    \includegraphics[width=.5\textwidth]{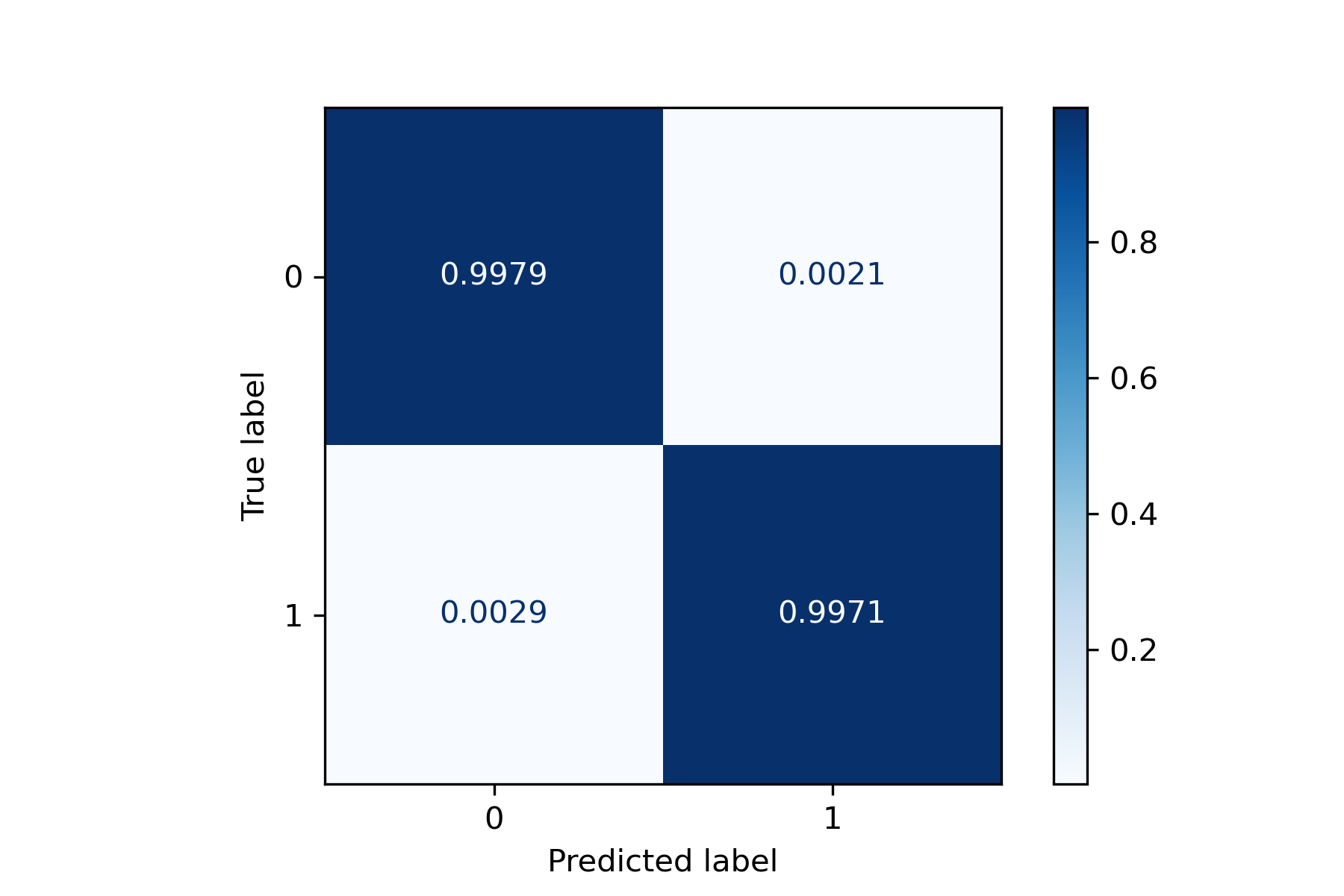}
    \caption{Trained model detection results without adversarial samples}
    \label{fig:bot1}
\end{figure}

Figure \ref{fig:bot1} shows the confusion matrix plot of the trained botnet detection system before applying the adversarial samples. The figure shows a FP rate of 0.21\%, and a FN rate of 0.29\% only. This shows the high accuracy of  the trained system.

In the next step of the experiment, we calculated SHAP values for the trained classifier using the dataset (including benign and malicious samples). The calculations are performed using the trained model, and the dataset input values, without using the dataset labels. This is aligned with our proposed attack's threat model that is based on a blackbox attack.
The explainer used in the attack was tree explainer due to the model being XGB-based. 

SHAP values calculate the impact of each feature on each prediction for all samples used in the calculation. In our experiment, we utilized the while dataset with 919,920 samples to get the highest possible accuracy in SHAP calculations.

\begin{figure*}
    \centering
    \includegraphics[width=.8\textwidth]{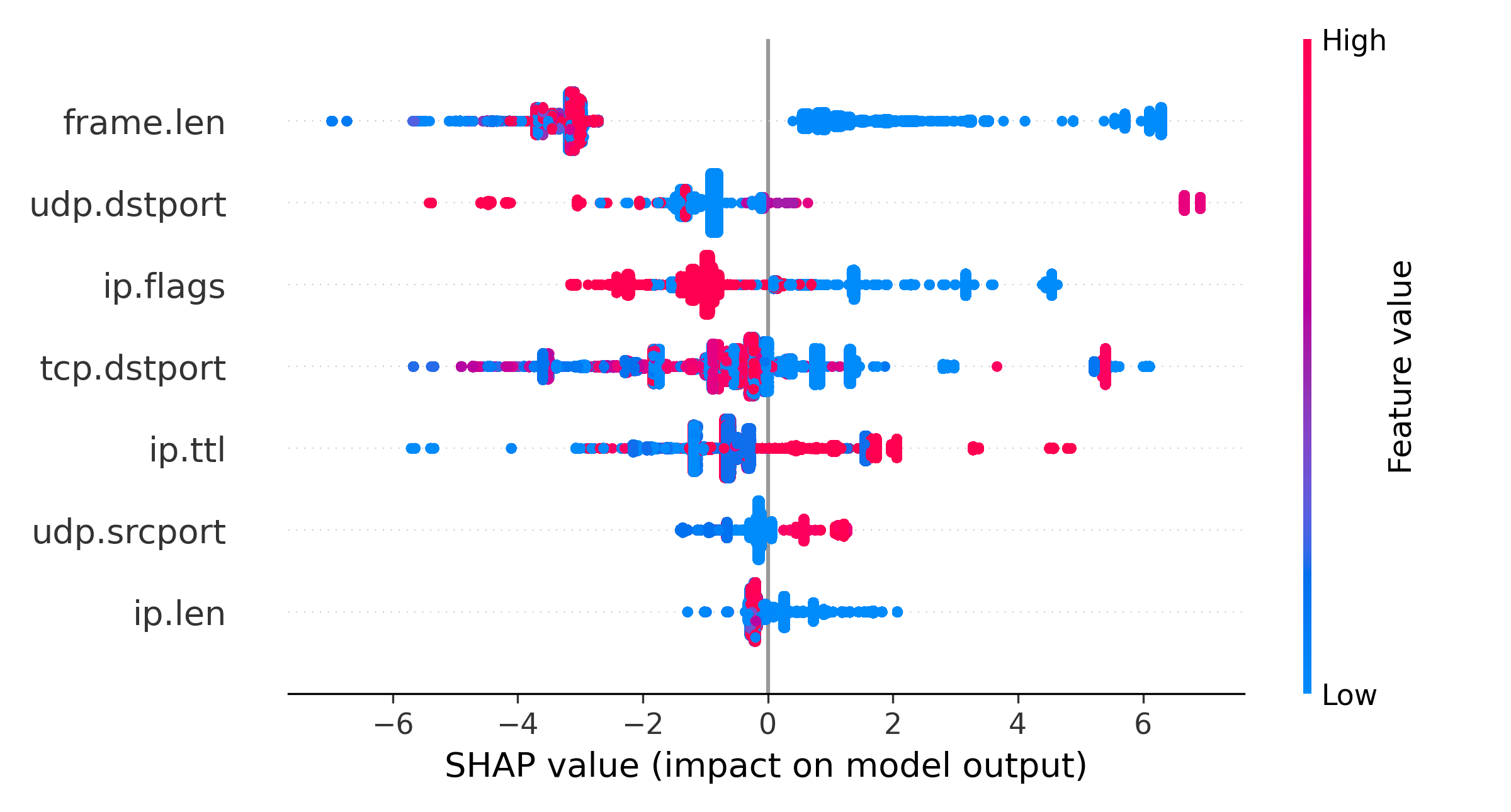}
    \caption{SHAP summary plot for botnet detection system}
    \label{fig:bot-shap}
\end{figure*}

The summary of the results of the SHAP calculations is shown in Figure \ref{fig:bot-shap}. Within this figure, each dot represents one data point, or in our scenario one packet. Dots on the right side of the axis represent points where the feature's value pushes the prediction to higher value (i.e. malicious). Dots on the left side of the axis are points where the feature's value pushes the prediction to a lower value (i.e. benign). The blue color represents low feature value, while the red color represents high feature value.

In Figure \ref{fig:bot-shap}, the features are ordered according to thier impact on the decision where the first feature has the highest impact, and the last feature has least impact on the decision. As shown, the \texttt{frame.len} feature holds the higest impact on the predicted decision, according to SHAP values. It is shown that the lowest values of \texttt{frame.len} push the prediction towards ``malicious''. This is aligned with the fact that most attack packets generated by botnets are either small scanning packets, or short well-crafted packets that try to break into an adjacent device~\cite{ref23}. Even when the botnet malware is communicating with the command and control (C2) center, it send small-sized ``keep-alive'' messages. On the other side, the figure shows that high values of the \texttt{frame.len} feature (represented by the red dots) would push the prediction closer to ``benign''.

In addition to this figure, the feature with the highest impact can be found be calculating the average of the absolute SHAP values for each feature across all samples. This produces an ``importance'' value attached to each feature. The results of the calculations is shown in Table \ref{tab:shaps}.

\begin{table}
    \centering
    \caption{Individual feature average SHAP value}
    \begin{tabular}{lc}
        \hline
        \textbf{Feature} & \textbf{Average SHAP value}\\
        \hline
        \texttt{frame.len} & 3.433418 \\
        \texttt{udp.dstport} & 2.018903 \\
        \texttt{ip.flags} & 1.633189 \\
        \texttt{tcp.dstport} & 1.386476 \\
        \texttt{ip.ttl} & 1.047884 \\
        \texttt{udp.srcport} & 0.429822 \\
        \texttt{ip.len} & 0.283653 \\
        \hline      
    \end{tabular} \label{tab:shaps}
\end{table}

As shown in the table, \texttt{frame.len} feature has the highest impact on the prediction decision.

The next step in the proposed attack is to find the best value of the \texttt{frame.len} that can be used to generate adversarial samples that could trick the detection system into classifying them as ``benign''. In the proposed attack, this process starts by searching for the lowest SHAP value for the selected feature (\texttt{frame.len} in the current experiment). The reason for searching for the lowest value is that the lowest SHAP, which would usually be a negative amount, is where the value of the feature is causing the largest impact in pushing the prediction to a lower value (zero, i.e. benign). On the other hand, the highest SHAP value would represent the feature value where the feature caused the largest push towards a high prediction (one, i.e. malicious).

\begin{figure*}
    \centering
    \includegraphics[width=.75\textwidth]{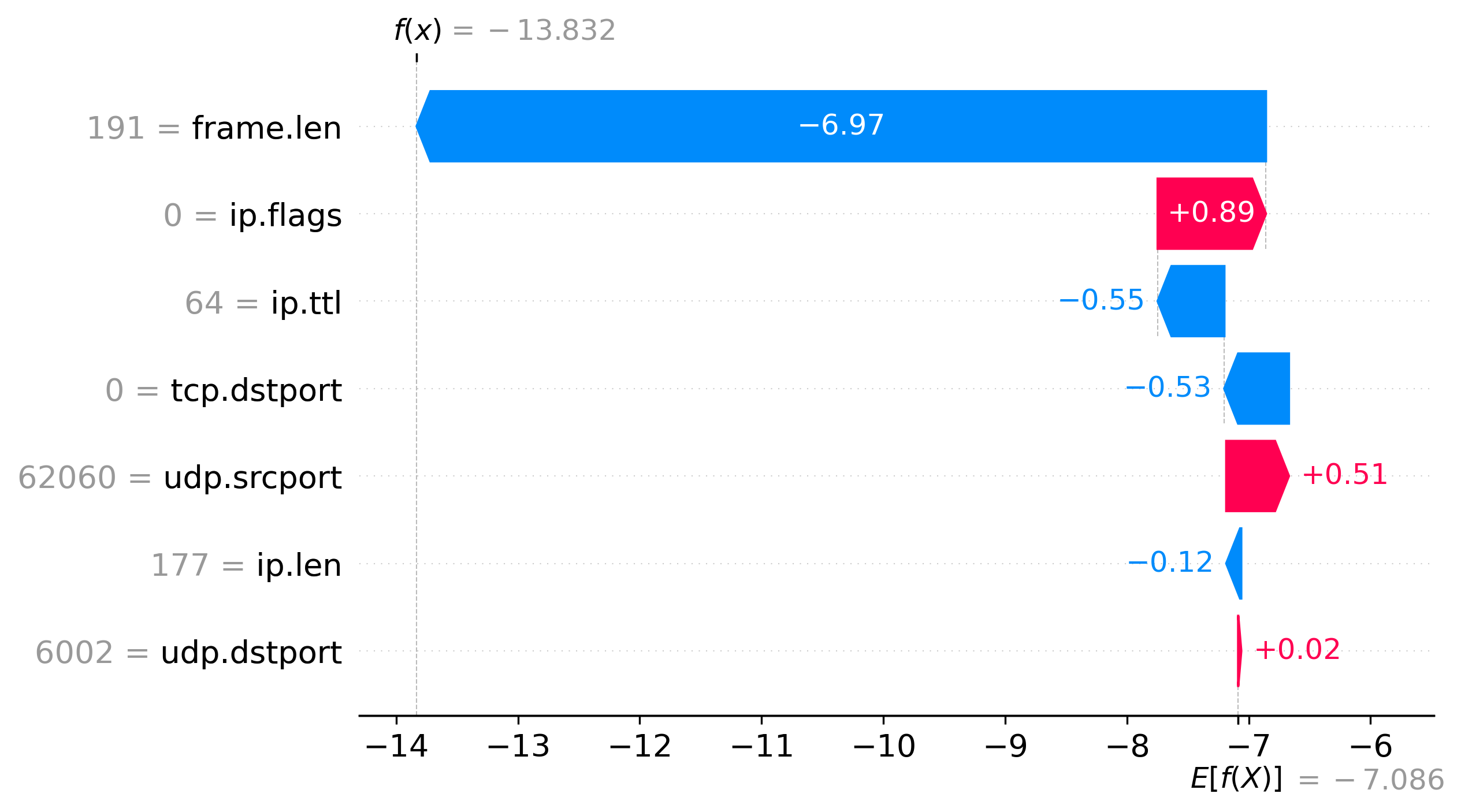}
    \caption{Waterfall plot for SHAP values of the sample with the lowest SHAP value in the \texttt{frame.len} feature}
    \label{fig:best}
\end{figure*}

The algorithm found that the lowest SHAP value for \texttt{frame.len} was found to be $-6.79$, which was associated with \texttt{frame.len} value of 191. Figure \ref{fig:best} shows the waterfall plot for the SHAP values of the sample with the lowest SHAP value in the \texttt{frame.len} feature. The figure shows the impact of the value of each feature on the prediction decision of one sample only. The particular sample shown is the one that had the lowest SHAP value for the \texttt{frame.len} feature. By default, SHAP explains XGB classifiers in term of their margin output, before the logistic link function. Therefore, the units of the x-axis in the figure are log-odds unit. This means that the negative value implies probabilities of less than 0.5, and hence leading the prediction to be zero (benign)~\cite{ref22}.

As shown in Figure \ref{fig:best}, the impact of the value 191 in \texttt{frame.len} was very significant in pushing the prediction to the left side. Therefore, we will be using this value to generate the adversarial samples for our attack.

In the next step of the attack, we generate adversarial samples by replacing the \texttt{frame.len} value in all malicious samples, all 228,251 of them, with 191. We kept the ``malicious'' label on them to measure the detection accuracy in identifying them as malicious. 

Once the adversarial samples were generated, they were used to test the trained detection model. During this test, the trained botnet detection model that previously demonstrated accuracy exceeding 99\% in detecting malicious samples failed to detect any of these samples. As shown in Table \ref{tab:advex-results}, all of the 228,251 adversarial samples were classified as benign.

\begin{table}
    \centering
    \caption{Testing results after creating the adversarial samples}
    \begin{tabular}{rlll}
        \hline
        & \textbf{Precision} & \textbf{Recall} & \textbf{$F_1$ Score}\\
        \hline
        \textbf{0} & 0.7515 & 0.9979 & 0.8574 \\
        \textbf{1} & 0.0000 & 0.0000 & 0.0000 \\
        \textbf{Accuracy} &  &  & 0.7503 \\
        \textbf{Macro average} & 0.3757 & 0.4990 & 0.4287 \\
        \textbf{Weighted average} & 0.5650 & 0.7503 & 0.6446 \\
        \hline        
    \end{tabular} \label{tab:advex-results}
\end{table}

\begin{figure}
    \centering
    \includegraphics[width=.50\textwidth]{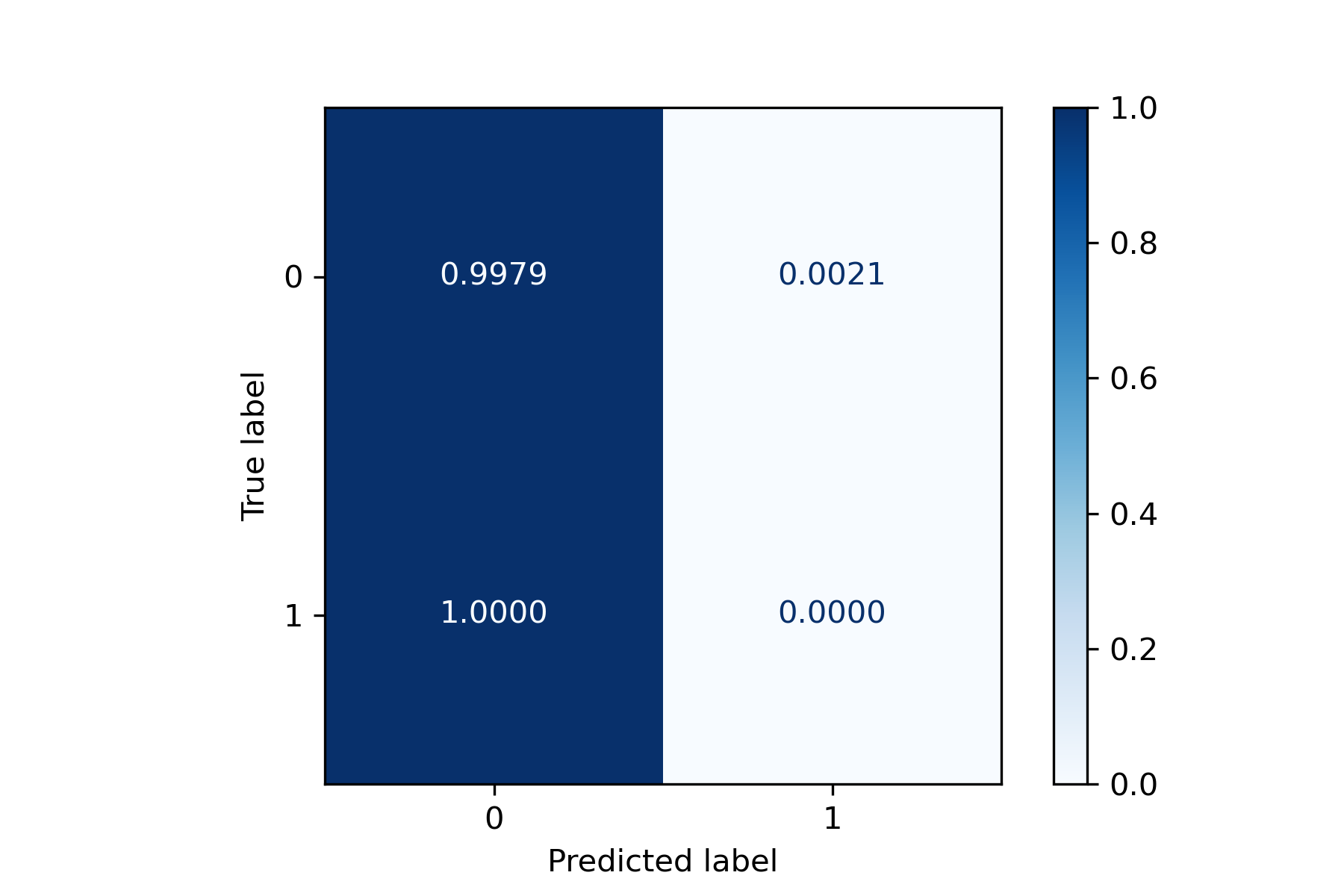}
    \caption{Trained model detection results with adversarial samples}
    \label{fig:bot2}
\end{figure}

As demonstrated in the table, none of the adversarial samples was classified as ``malicious''. All traffic was classified as benign. Figure \ref{fig:bot2} shows the confusion matrix plot resulting from testing the botnet detection classifier with normal and adversarial samples. While the detector delivered a FP rate of 0.21\%, it delivered a FN rate of 100\%. This indicates that the system failed in detecting any of the adversarial samples giving the attacker the ability to fully evade the detection system with 100\% success by applying perturbations to one feature only.

\section{DISCUSSIONS} \label{sec:discussions}
The goal of this research is to propose an adversarial attack that can utilize model's explainability in generating adversarial samples that can evade detection. The implementation and testing of the proposed attack has proved that the detection model's explainability, even in a black box scenario, can be used to provide the attacker with a deep understanding of how the detection process works in the trained model.

The proposed attack succeeded in identifying the most effective feature in a trained model, and find the required perturbation from the model's explainability calculations obtained from SHAP values. Tests showed that the proposed attack has achieved its design goal by succeeding in converting all malicious in the dataset (228,251 samples) to adversarial samples, and showed that these adversarial samples evaded detection with a 100\% success rate.

In the following subsections we will compare the proposed attack with related works, and discuss practical implementation considerations.

\subsection{Comparative Analysis}
Table \ref{tab:compare} shows a detailed comparison of the proposed attack to a group of related works. As the number of works addressing detection evasion of botnets, we have included a few works that address intrusion detection evasion.

\begin{table*}
    \centering
    \caption{Comparison of the proposed attack to related works}
    \begin{adjustbox}{width=1\textwidth,center=\textwidth}
    \begin{tabular}{lp{2.8cm}p{3cm}cccc}
        \hline
        \textbf{Work} & \textbf{Technique} & \textbf{Dataset} & \textbf{Feature(s)} & \textbf{Success \%} & \textbf{Focus} & \textbf{Blackbox}\\
        \hline
        \cite{ref15} & JSMA & CIC-IDS-2017 & - & 22 & IDS & \xmark\\
        \cite{ref16} & Adaptive-JSMA & NSL-KDD & 5 & 95 & IDS & \xmark\\
        \cite{ref11} & JSMA & CIC-IDS-2017 & 3 & 18 & IDS & \cmark\\
        \cite{ref26} & Mean benign values & CTU-13 and CSE-CIC-IDS2018 & 8 & 78 to 95 & Botnet & \cmark \\
        Proposed & Adversarial Explainability & IoT-ID & 1 & 100 & Botnet & \cmark \\
        \hline
    \end{tabular} \label{tab:compare}
    \end{adjustbox}
\end{table*}

As shown in the table, most adversarial attacks are directed towards intrusion detection systems. These attacks rely on JSMA to find the most effective group of features. JSMA is an attack focused on creating adversarial samples through the introduction of noise to the data to cause the ML-based classifier to misclassify the input. The attack was originally introduced to target image classification systems. However, it was later expanded to be implemented on many other types of classifiers. Throughout the years, many variations of the original attack were proposed such as \cite{ref29}. The attack proposed in \cite{ref26} did not use JSMA or any of its variants. Instead, the attack utilized mean calculation of the benign sample values and used them to generate perturbation values in an iterative manner, where the perturbation value starts small, and is increased repeatedly.

While the attack proposed in \cite{ref15} did not state the number of features altered, the attacks proposed in \cite{ref16, ref11, ref26} required perturbations to be applied to a larger number of features (3-8). In comparison, our proposed attack requires the change of one feature only. In addition, the methodologies used in these four attacks require iterative operations that keep changing the perturbations until a successful value is reached. Our proposed method finds the exact value that would cause detection evasion without any trial and error attempts.

As shown in Table \ref{tab:compare}, our proposed attack achieved 100\% success, while other attacks could not exceed 95\% using a larger number of feature perturbations. This, in addition to the points mentioned previously, establishes that our proposed attack outperforms other adversarial attacks focused on detection evasion.

\subsection{Practical Implementation Challenges}
While our experiments have shown that the proposed attack is effective in evading machine learning-based botnet detection, there are several points to discuss regarding the implementability of the attack in real-life scenarios.

The nature of the most effective feature has a significant impact on whether the attack can be implemented in real-life scenarios or not. In the model used to demonstrate the attack, \texttt{frame.len} feature was the most effective one. In real-life attacks, this feature can be easily manipulated to implement the perturbations chosen. It is possible to inject empty padding bits, for instance, without impacting the actual frame contents. However, there are many other features that are can pose a challenge and cannot be easily manipulated. An example of such feature can be \texttt{tcp.dstport}. While the source port number can easily be manipulated by the attacker that is originating the traffic, it might not be possible for the attacker to change the destination port number for the attack, as this is decided by the target, not the attacker. In a scenario where the most effective feature might not be practically altered, the attacker can try using the second most effective feature to evade detection. While perturbations in the second most effective feature might not be as effective as the first, it is worth the try by the attacker.

Another factor that might impact the practicality of the attack is the most effective value of the selected feature. There might be scenarios where the most effective feature is one that is controlled by the attacker, but the selected value might be difficult to achieve. In a scenario where the most effective feature could be \texttt{frame.len}, and the most effective value being a small value, such as 50. It might not always be possible for the attacker to craft the attack packets to such a small size. While many higher level data can be subjected to IP fragmentation, it might nmot be possible all the time.

\section{CONCLUSIONS} \label{sec:conclusions}
In this paper, we presented a novel adversarial attack concept that employs models' explainability in bypassing cybersecurity detection systems that are based on machine learning, named \textit{adversarial explainability}. The proposed attack utilizes perturbations in the features identified as the most effective features by the model's explainability. The creation of the adversarial samples then utilizes explainability once more in finding the most effective value of the perturbation to help the attacker bypass ML-based botnet detection.

The proposed attack was tested on a trained botnet detection classifier, and was successful in altering 228,251 malicious samples to generate adversarial samples. The generated adversarial samples were 100\% successful in bypassing detection.

Future research directions include exploring the implementation of the proposed attack concept on various cybersecurity detection systems, such as intrusion detection, and malware detection, to measure the attack performance. Another research direction would be exploring methods to counter this attack and protect ML-based detection systems from it.

\section{ACKNOWLEDGMENTS}
This work was funded by NSERC, Canada.

\bibliographystyle{ieeetr}
\bibliography{references}

\begin{thebibliography}{10}

\bibitem{ref1}
``{Artificial Intelligence market size/revenue comparisons 2022 {$\vert$}
  Statista},'' Mar. 2023.
\newblock [Online; accessed 23. Mar. 2023],
  https://www.statista.com/statistics/941835/artificial-intelligence-market-size-revenue-comparisons.

\bibitem{ref2}
M.~M. Alani, ``Big data in cybersecurity: a survey of applications and future
  trends,'' {\em Journal of Reliable Intelligent Environments}, vol.~7, no.~2,
  pp.~85--114, 2021.

\bibitem{ref4}
I.~H. Sarker, A.~Kayes, S.~Badsha, H.~Alqahtani, P.~Watters, and A.~Ng,
  ``Cybersecurity data science: an overview from machine learning
  perspective,'' {\em Journal of Big data}, vol.~7, pp.~1--29, 2020.

\bibitem{ref23}
M.~M. Alani, ``Botstop: Packet-based efficient and explainable iot botnet
  detection using machine learning,'' {\em Computer Communications}, vol.~193,
  pp.~53--62, 2022.

\bibitem{ref24}
Y.~Xing, H.~Shu, H.~Zhao, D.~Li, and L.~Guo, ``Survey on botnet detection
  techniques: Classification, methods, and evaluation,'' {\em Mathematical
  Problems in Engineering}, vol.~2021, pp.~1--24, 2021.

\bibitem{ref25}
S.~Gaonkar, N.~F. Dessai, J.~Costa, A.~Borkar, S.~Aswale, and P.~Shetgaonkar,
  ``A survey on botnet detection techniques,'' in {\em 2020 International
  Conference on Emerging Trends in Information Technology and Engineering
  (ic-ETITE)}, pp.~1--6, IEEE, 2020.

\bibitem{ref10}
M.~M. Alani, ``On recent security issues in machine learning,'' in {\em 2020
  International Conference on Software, Telecommunications and Computer
  Networks (SoftCOM)}, pp.~1--6, IEEE, 2020.

\bibitem{ref13}
E.~Alhajjar, P.~Maxwell, and N.~Bastian, ``Adversarial machine learning in
  network intrusion detection systems,'' {\em Expert Systems with
  Applications}, vol.~186, p.~115782, 2021.

\bibitem{ref14}
Z.~Zhang, H.~A. Hamadi, E.~Damiani, C.~Y. Yeun, and F.~Taher, ``Explainable
  artificial intelligence applications in cyber security: State-of-the-art in
  research,'' {\em IEEE Access}, vol.~10, pp.~93104--93139, 2022.

\bibitem{ref15}
Z.~Lin, Y.~Shi, and Z.~Xue, ``Idsgan: Generative adversarial networks for
  attack generation against intrusion detection,'' in {\em Advances in
  Knowledge Discovery and Data Mining: 26th Pacific-Asia Conference, PAKDD
  2022, Chengdu, China, May 16--19, 2022, Proceedings, Part III}, pp.~79--91,
  Springer, 2022.

\bibitem{ref12}
M.~A. Ayub, W.~A. Johnson, D.~A. Talbert, and A.~Siraj, ``Model evasion attack
  on intrusion detection systems using adversarial machine learning,'' in {\em
  2020 54th annual conference on information sciences and systems (CISS)},
  pp.~1--6, IEEE, 2020.

\bibitem{ref16}
R.~Sheatsley, N.~Papernot, M.~J. Weisman, G.~Verma, and P.~McDaniel,
  ``Adversarial examples for network intrusion detection systems,'' {\em
  Journal of Computer Security}, vol.~2022, no.~Preprint, pp.~1--26, 2022.

\bibitem{ref11}
H.~Mohammadian, A.~A. Ghorbani, and A.~H. Lashkari, ``A gradient-based approach
  for adversarial attack on deep learning-based network intrusion detection
  systems,'' {\em Applied Soft Computing}, p.~110173, 2023.

\bibitem{ref26}
I.~Debicha, B.~Cochez, T.~Kenaza, T.~Debatty, J.-M. Dricot, and W.~Mees,
  ``Adv-bot: Realistic adversarial botnet attacks against network intrusion
  detection systems,'' {\em Computers \& Security}, vol.~129, p.~103176, 2023.

\bibitem{ref18}
C.~Zhang, X.~Costa-Pérez, and P.~Patras, ``Adversarial attacks against deep
  learning-based network intrusion detection systems and defense mechanisms,''
  {\em IEEE/ACM Transactions on Networking}, vol.~30, no.~3, pp.~1294--1311,
  2022.

\bibitem{ref22}
C.~Molnar, {\em Interpretable machine learning}.
\newblock Lulu. com, 2020.

\bibitem{ref19}
S.~M. Lundberg and S.-I. Lee, ``A unified approach to interpreting model
  predictions,'' {\em Advances in neural information processing systems},
  vol.~30, 2017.

\bibitem{ref28}
M.~Mohammed, M.~B. Khan, and E.~B.~M. Bashier, {\em Machine learning:
  algorithms and applications}.
\newblock Crc Press, 2016.

\bibitem{ref27}
H.~Kang, D.~H. Ahn, G.~M. Lee, J.~D. Yoo, K.~H. Park, and H.~K. Kim, ``Iot
  network intrusion dataset,'' 2019.

\bibitem{ref29}
R.~Wiyatno and A.~Xu, ``Maximal jacobian-based saliency map attack,'' {\em
  arXiv preprint arXiv:1808.07945}, 2018.

\end{thebibliography}

\begin{IEEEbiography}{Mohammed M. Alani}{\,}is a Research Fellow at the Cybersecurity Research Lab, toronto Metropolitan University. He received his PhD in 2007. He is a senior member of the IEEE and the ACM. His current research interests include IoT security, intrusion detection, and malware analysis and detection. Contact him at m@alani.me
\end{IEEEbiography}

\begin{IEEEbiography}{Atefeh Mashatan}{\,} is a Canada Research Chair and Associate Professor at Ted Rogers School of Information Technology Management. Collaborating with industry partners, Dr. Mashatan studies industry relevant research problems and proposes solutions that can be developed as part of the industry-academic collaborations. Her expertise at the frontlines of the global cybersecurity field was recognized by SC Magazine in 2019, when she was named one of the top five Women of Influence in Security. In 2020, she received the Enterprise Blockchain Award in the category of New Frontiers in Blockchain Academic Research by Blockchain Research Institute.\vspace*{8pt}
\end{IEEEbiography}

\begin{IEEEbiography}{Ali Miri} {\,}is a Full Professor at the School of Computer Science, Ryerson University, Toronto. He has over 25 years of research experience in security and privacy technologies and their applications, computer networks and digital communication, and cloud computing and big data. He has authored and co-authored over 240 refereed manuscripts. Dr. Miri has served on more than 100 organizing and technical program committees of international conferences and workshops, and has been the main organizer of over a dozen international conferences. He is a senior member of the IEEE, and a member of the Professional Engineers of Ontario.
\end{IEEEbiography}

\end{document}